\documentclass[aps,prl,oneside,twocolumn]{revtex4-1}
\usepackage{amsmath}
\usepackage{amssymb}
\usepackage{graphicx}

\begin{document}

\draft \title{Lowest-order relativistic corrections to the fundamental limits of nonlinear-optical coefficients}
\author{Nathan J. Dawson}
\address{Department of physics, Case Western Reserve University, Cleveland, OH 44106, USA}
\email{dawsphys@hotmail.com}
\date{\today}

\begin{abstract}The effects of small relativistic corrections to the off-resonant polarizability, hyperpolarizability, and second hyperpolarizability are investigated. Corrections to linear and nonlinear optical coefficients are demonstrated in the three-level ansatz, which includes corrections to the Kuzyk limits when scaled to semi-relativistic energies. It is also shown that the maximum value of the hyperpolarizability is more sensitive than the maximum polarizability or second hyperpolarizability to lowest-order relativistic corrections. These corrections illustrate how the intrinsic nonlinear-optical response is affected at semi-relativistic energies.\end{abstract}


\maketitle

\section{Introduction}

Over a decade ago, Kuzyk \cite{kuzyk00.01} showed that there are fundamental limits to the off-resonant, electronic, nonlinear-optical response. This was discovered by manipulating both the on- and off-diagonal elements of the Thomas-Reiche-Kuhn (TRK) sum rule,\cite{thoma25.01,reich25.01,kuhn25.01} which limits the oscillator strengths of a quantum system with respect to fundamental constants in the non-relativistic regime. The oscillator strength is limited by the non-relativistic kinetic energy of a free particle, where field interactions from a four-potential do not contribute to the maximum oscillator strength. The intrinsic values of the hyperpolarizability and second hyperpolarizability in the non-relativistic limit have been studied in great detail,\cite{kuzyk00.02,kuzyk01.01,clays01.01,kuzyk03.03,clays03.01,Tripa04.01,perez05.01,kuzyk05.02,kuzyk06.01,kuzyk06.03,perez07.01,perez07.02,perez01.08,zhou08.01,kuzyk09.01,dawson11.02,dawson11.03,shafe13.01,kuzyk13.01} where there is a looming gap between the measured/calculated values and the fundamental limits in the non-relativistic regime.

There have been several approaches to reduce this gap using optimization routines on one-dimensional potentials, which have resulted in the confirmation of the apparent gap.\cite{zhou06.01,zhou07.02,wigge07.01,kuzyk08.01,watkins09.01,shafe10.01,watkins11.01,ather12.01,burke13.01} Another approach to breach the gap involves a systematic search for new classes of organic nonlinear optical molecules with multipolar charge-density analysis from crystallographic data.\cite{cole02.01,cole03.01,higgi12.01} New abstract methods of calculating large nonlinear responses have also been studied for low-dimensional quantum graphs.\cite{shafe12.01,lytel13.01,lytel13.02} All of these approaches focus on breaching the gap between the fundamental limit and the largest calculated (or directly measured) intrinsic values.

Although a four-potential does not contribute to the non-relativistic TRK sum rule, there may be other ways to change the limiting value on the oscillator strength, and thereby adjust the fundamental limits of the nonlinear optical response. Instead of focusing on optimizing the intrinsic value based on a specific potential, I will discuss the changes in the limiting constant of the TRK sum rules for a specific type of quantum systems that are not properly represented by a closed Schrodinger equation. Specifically, a relativistic system is examined which no longer has a simple $p^2/2m$ kinetic energy approximation, and therefore, the energy-momentum relationship directly affects the fundamental limits. Thus, this paper is dedicated to the study of the fundamental limits of the hyperpolarizability and second hyperpolarizability for systems that have non-negligible relativistic energies.



\section{Theory}

In the far off-resonant limit (frequency approaches zero), the respective one-dimensional polarizability, hyperpolarizability, and second hyperpolarizability are \cite{orr71.01}
\begin{eqnarray}
\alpha &=& 2 e^2 \left. \displaystyle \sum_{n}^{\infty} \right.^{\prime} \frac{x_{0n} x_{n0}} {E_{n0}} , \label{eq:polar} \\
\beta &=& 3 e^3 \left. \displaystyle \sum_{n,l}^{\infty} \right.^{\prime} \frac{x_{0n} \bar{x}_{nl} x_{l0}} {E_{n0} E_{l0}} , \label{eq:hyper}
\end{eqnarray}
and
\begin{equation}
\gamma = 4 e^4 \left( \left.\displaystyle \sum_{n,l,k}^{\infty} \right.^{\prime} \frac{x_{0n} \bar{x}_{nl} \bar{x}_{lk} x_{k0}} {E_{n0} E_{l0} E_{k0}} - \left.\displaystyle \sum_{n,l}^{\infty} \right.^{\prime} \frac{x_{0n} x_{n0} x_{0l} x_{l0}} {E_{n0}^2 E_{l0}} \right) ,
\label{eq:sechyper}
\end{equation}
where $x$ is the position operator in one-dimension, $e$ is the magnitude of an electron's charge, $E_{i}$ is the $i$th energy eigenstate, and the prime restricts the summation by excluding the ground state. The shorthand notation, $x_{ij} = \left\langle i \left| x \right| j \right\rangle$ and $E_{ij} = E_{i} - E_{j}$, was introduced in Eqs. \ref{eq:polar}-\ref{eq:sechyper}. Note that the barred operator presented in the expressions for the nonlinear coefficients is the origin-specific expectation value, which is given as $\bar{x}_{ii} = x_{ii} - x_{00}$ when the indices are matched.

The TRK sum rules for the Dirac equation, gives the well-known result of all states summing to zero, where the zero value is due to the sum of the positive and corresponding negative energy states.\cite{levin57.01} We wish to only observe the positive energy states from an electron in an atom or molecule, and therefore, we must project out the positive energy states. For a single electron system, the \textit{positive energy} TRK sum rules to lowest-relativistic order (ordered in $1/c$) have previously been derived \cite{leung86.01,cohen98.01,sinky06.01} using a Foldy-Wouthuysen (FW) transformation,\cite{foldy50.01}
\begin{eqnarray}
& & \displaystyle\sum_{l = 0}^{\infty} \left\langle k \left| \boldsymbol{r} \right| l \right\rangle \left\langle l \left| \boldsymbol{r} \right| n \right\rangle \left[E_l - \frac{1}{2}\left(E_k + E_n\right)\right] \nonumber \\
&=& \left\langle k' \left|\left(\frac{3\hbar^2}{2m} + \frac{5\hbar^4}{4m^3 c^2}\nabla^2\right)\right| n' \right\rangle ,
\label{eq:3DrelTRK}
\end{eqnarray}
where $\left| n' \right\rangle = e^{i{\cal S}} \left| n \right\rangle$ with ${\cal S}$ being a unitary operator. Equation \ref{eq:3DrelTRK} differs slightly from the results of Ref. \cite{cohen98.01}, where we have derived the relation with arbitrary eigenstates because both the on- and off-diagonal components of the sum rules are essential in determining the off-resonant, nonlinear-optical responses.\cite{kuzyk00.01} In the FW approach $p = p'$, where an operator, $A$, in the FW approximation is defined as $A' = e^{i{\cal S}} A e^{-i{\cal S}}$. Thus, the momentum operator commutes with $e^{iS}$, and therefore $\left\langle k' \left| p^2 \right| n' \right\rangle = \left\langle k \left| e^{-i{\cal S}} p^2 e^{i{\cal S}} \right|n \right\rangle = \left\langle k \left| p^2 \right|n \right\rangle$. Note that while transforming the Hamiltonian for an electron interacting with fields, ${\cal S}$ is chosen at every iteration to remove all odd operators.

Inasmuch as the $\nabla$ operator is related to the momentum operator, and that there is an equivalence between the momentum operator in the Schrodinger equation and the momentum in the FW transformation, the right-hand-side (RHS) of Eq. \ref{eq:3DrelTRK} may be written as
\begin{equation}
\left\langle k \left|\left(\frac{3\hbar^2}{2m} - \frac{5 \hbar^2 p^2}{4m^3 c^2}\right)\right| n \right\rangle . \nonumber
\label{eq:RHSofTRK}
\end{equation}
Thus, to the lowest-order relativistic correction, the RHS of the TRK sum rules given in Eq. \ref{eq:3DrelTRK} decreases for any real value of the momentum.

The lowest-order relativistic approximation to the Hamiltonian (for an electron in the presence of a scalar potential only) is given as
\begin{eqnarray}
H &=& H_{0} - \frac{p^4}{8m^3 c^2} + \frac{1}{4 m^2 c^2} \left(\boldsymbol{\sigma}\cdot \boldsymbol{p}\right) V\left(\boldsymbol{r}\right)\, \left(\boldsymbol{\sigma}\cdot \boldsymbol{p}\right) \nonumber \\
&-& \frac{1}{8 m^2 c^2} \left(p^2 V\left(\boldsymbol{r}\right) + V\left(\boldsymbol{r}\right)\,p^2 \right) , \label{eq:lowestHbefore}
\end{eqnarray}
where
\begin{equation}
H_{0} = \frac{p^2}{2m} + V\left(\boldsymbol{r}\right)
\label{eq:Hschrod}
\end{equation}
with $V\left(\boldsymbol{r}\right)$ denoting a spatially dependent scalar potential and $\boldsymbol{\sigma}$ representing the Pauli spin matrices. We may rewrite Eq. \ref{eq:lowestHbefore} as the well-known result \cite{grein00.01}
\begin{eqnarray}
H &=& \frac{p^2}{2m} - \frac{p^4}{8m^3 c^2} + V\left(\boldsymbol{r}\right) + \frac{\hbar}{4 m^2 c^2}\, \boldsymbol{\sigma}\cdot \left\{\left[\nabla V\left(\boldsymbol{r}\right)\right]\times \boldsymbol{p}\right\} \nonumber \\
&+& \frac{\hbar^2}{8 m^2 c^2} \nabla^2 V\left(\boldsymbol{r}\right) . \label{eq:lowestH}
\end{eqnarray}
The Hamiltonian with lowest-order relativistic corrections is a quartic equation with respect to momentum. For central potentials, the spin-orbit term may be recast in terms of the angular momentum operator, and thereby reduces the Hamiltonian to a quadratic equation in $p^2$.

There is an alternative method of reducing Eq. \ref{eq:lowestH} to a quadratic that does not require one to collapse the parameter space to the centrosymmetric limit, which is observed when limiting the system to one-dimension. In one-dimension, there is no orbital angular momentum, $\nabla V \times \boldsymbol{p} = 0$, and therefore, the spin-orbit term vanishes. 
Thus, Eq. \ref{eq:lowestH} reduces to a simplified quadratic equation in $p_{x}^2$, where
\begin{eqnarray}
H = \frac{p_{x}^2}{2m} - \frac{p_{x}^4}{8m^3 c^2} + V\left(x\right)
+ \frac{\hbar^2}{8 m^2 c^2} \nabla^2 V\left(x\right) . \label{eq:lowestH1D}
\end{eqnarray}
Note that the Darwin still term survives the one-dimensional approximation. Although this approach simplifies the study of generalized semi-relativistic interactions while maintaining a non-centrosymmetric parameter space, one should note that many recent advances in quantum chemistry have been introduced for numerically approximating specified relativistic systems. Most notably are the electrostatic-potential-ordered Douglass-Kroll-Hess method,\cite{dougl74.01,hess86.01,nakaj00.01,reihe12.01} the ordered regular approximations,\cite{lenth93.01,lenth94.01,lenth96.01,filat03.01} and others based on exact decoupling methods.\cite{filat03.02,kutze05.01,kutze06.01,kutze07.01}

By restricting ourselves to one dimension, we may write
\begin{equation}
\left\langle k \left| V + \frac{p_{x}^2}{2m} - \frac{p_{x}^4}{8m^3 c^2} + \frac{\hbar^2}{8 m^2 c^2} \frac{\partial^2 V}{\partial x^2} = E_n \right| n \right\rangle . \label{eq:simpleham}
\end{equation}
Solving Eq. \ref{eq:simpleham} for $p_{x}^2$ gives
\begin{eqnarray}
\left\langle k \left| p_{x}^{2} \right| n \right\rangle &=& 2m^2c^2\delta_{k,n} - 2m^2c^2 \label{eq:12345} \\
&\times& \left\langle k \left|\sqrt{1 - \frac{2 \left(E_n- V\right)}{ m c^2} + \frac{\hbar^2}{4 m^3 c^4}\frac{\partial^2 V}{\partial x^2} } \right| n \right\rangle , \nonumber
\end{eqnarray}
where $\delta$ is the Kronecker delta function, and the negative root was chosen which reduces Eq. \ref{eq:12345} to the non-relativistic TRK sum rules as $1/c \rightarrow 0$. Using Eq. \ref{eq:12345}, the one-dimensional TRK sum rule with lowest-order relativistic corrections,
\begin{eqnarray}
& & \displaystyle\sum_{l = 0}^{\infty} \left\langle k \left| x \right| l \right\rangle \left\langle l \left| x \right| n \right\rangle \left[E_l - \frac{1}{2}\left(E_k + E_n\right)\right] \nonumber \\
&=& \left\langle k \left|\left(\frac{\hbar^2}{2m} - \frac{3 \hbar^2 p_{x}^2}{4m^3 c^2}\right)\right| n \right\rangle ,
\label{eq:1DrelTRK}
\end{eqnarray}
may be rewritten as
\begin{eqnarray}
& & \displaystyle\sum_{l = 0}^{\infty} \left\langle k \left| x \right| l \right\rangle \left\langle l \left| x \right| n \right\rangle \left[E_l - \frac{1}{2}\left(E_k + E_n\right)\right] \nonumber \\
&=& \frac{\hbar^2}{m}\left(\frac{3}{2}\lambda_{kn} - \delta_{k,n}\right),
\label{eq:1DrelTRKsubbed}
\end{eqnarray}
where
\begin{equation}
\lambda_{kn} = \left\langle k \left|\sqrt{1 - \frac{2}{ m c^2} \left(E_n- V\right) + \frac{\hbar^2}{4 m^3 c^4}\frac{\partial^2 V}{\partial x^2} } \right| n \right\rangle .
\label{eq:lambdakn}
\end{equation}

Under the current set of approximations, we take the element ($k=0$, $n=0$), or (0,0), which gives
\begin{equation}
\left|x_{10}\right|^2 E_{10} = \frac{\hbar^2}{m} \left(\frac{3}{2}\lambda_{00} - 1\right) - \displaystyle\sum_{l=2}^{\infty} \left|x_{l0}\right|^2 E_{l0} .
\label{eq:00eqfull}
\end{equation}
Considering the diagonal components and neglecting the Darwin term, there are two regimes that adjust the fundamental limit. If $E_{n} > V_{n,n}\left(x\right)$, then the electron is moving inside a potential and $\lambda_{nn}$ is real. This causes a decrease in the maximum oscillator strength. If the electron is expected to be outside a potential such that $E_{n} < V_{n,n}\left(x\right)$, then $\lambda_{nn}$ becomes imaginary, which cannot occur for bound states with positive energies. Therefore, the oscillator strength of a one-dimensional semi-relativistic system decreases with respect to the non-relativistic approximation; however, a competing parameter may increase the final numerical value of some systems (not the intrinsic value) of the off-resonant response because the relativistic corrections reduce the transition energies with respect to those mapped from the non-relativistic Hamiltonian.

In prior studies that began with a Hamiltonian in the non-relativistic limit, it was shown that the largest nonlinear-optical responses occur when all other transition energies become much larger than $E_{10}$. In other words, the sum-over-state (SOS) expressions are dominated by the first excited state transition. This is also true for relativistically corrected systems which is obvious from Eq. \ref{eq:00eqfull}. Thus, we adopt the same method as Kuzyk \cite{kuzyk00.01} and assume a three-level model. Then, Eq. \ref{eq:00eqfull} reduces to
\begin{equation}
\left|x_{10}\right|^2 E_{10} + \left|x_{20}\right|^2 E_{20} = \frac{\hbar^2}{m} \left(\frac{3}{2}\lambda_{00} - 1\right) .
\label{eq:00eq}
\end{equation}
Likewise, (1,1) produces the resultant equation
\begin{equation}
\left|x_{12}\right|^2 E_{21} - \left|x_{10}\right|^2 E_{10} = \frac{\hbar^2}{m} \left(\frac{3}{2}\lambda_{11} - 1\right) .
\label{eq:11eq}
\end{equation}

In the same manner as Eqs. \ref{eq:00eq} and \ref{eq:11eq} we take (1,0), which gives
\begin{equation}
x_{10} \bar{x}_{11} E_{10} + x_{12}x_{20} \left(E_{21} + E_{20}\right) = \frac{3 \hbar^2}{2m} \lambda_{10},
\label{eq:10eq}
\end{equation}
Note that the left-hand-side of Eq. \ref{eq:10eq} is identical for (1,0) and (0,1) when we assume real transition moments, \textit{i}.\textit{e}., $x_{ij} = x_{ji}$.\cite{kuzyk01.01} Thus, it is of no surprise that the corresponding $\lambda$ parameter must also possess the property $\lambda_{10} = \lambda_{01}$. Finally, we set the matrix elements corresponding to (2,0), or (0,2), which gives
\begin{equation}
x_{20} \bar{x}_{22} E_{20} + x_{10}x_{12} \left(E_{10} - E_{21}\right) = \frac{3 \hbar^2}{2m} \lambda_{20} .
\label{eq:20eq}
\end{equation}
Note that Eqs. \ref{eq:10eq} and \ref{eq:20eq} contain off-diagonal components that are real and positive for well behaved systems. 

Solving Eqs. \ref{eq:00eq}-\ref{eq:20eq} for the transition dipole moments, we find
\begin{eqnarray}
\left| x_{10} \right| &=& \frac{\hbar}{\sqrt{m E_{10}}} X \sqrt{\frac{3}{2}\lambda_{00} - 1} , \label{eq:x10fs} \\
\left| x_{12} \right| &=& \frac{\hbar}{\sqrt{m E_{10}}} \sqrt{\frac{E}{1-E}} \, G_{\lambda}\left(X\right) , \label{eq:x12fs} \\
\bar{x}_{11} &=& \frac{\hbar}{\sqrt{m E_{10}}} \left[ \displaystyle \frac{E-2}{\displaystyle \sqrt{1-E}} \frac{\displaystyle \sqrt{1-X^2}}{X} \, G_{\lambda}\left(X\right) \right. \nonumber \\
&+& \left. \frac{3\lambda_{10}}{2X \sqrt{\frac{3}{2}\lambda_{00} - 1}} \right] , \label{eq:x11fs} \\
\bar{x}_{22} &=& \frac{\hbar}{\sqrt{m E_{10}}} \left[ \displaystyle \frac{1-2E}{\displaystyle \sqrt{1-E}} \frac{X}{\displaystyle \sqrt{1-X^2}} \, G_{\lambda}\left(X\right) \right. \nonumber \\
&+& \left. \frac{3 \sqrt{E} \lambda_{20}}{2 \displaystyle \sqrt{1-X^2} \sqrt{\frac{3}{2}\lambda_{00} - 1}} \right] , \label{eq:x22fs}
\end{eqnarray}
and
\begin{equation}
\left| x_{20} \right| = \frac{\hbar}{\sqrt{m E_{10}}}\sqrt{E} \sqrt{1-X^2} \sqrt{\frac{3}{2}\lambda_{00} - 1} .
\label{eq:x20fs}
\end{equation}
where
\begin{equation}
G_{\lambda}\left(X\right) = \sqrt{X^2 \left(\frac{3}{2}\lambda_{00} - 1\right) + \frac{3}{2}\lambda_{11} - 1} .
\label{eq:Glx}
\end{equation}
Here, we used the notation inline with previous expressions for the nonlinear-optical limits of non-relativistic systems such that
\begin{equation}
X = \frac{\left|x_{10}\right|}{\left|x_{10}^{\mathrm{max}}\right|} \qquad \mathrm{and} \qquad E = \frac{E_{10}}{E_{20}} ,
\label{eq:energyfrac}
\end{equation}
where we can see that the maximum value for the $x_{10}$ transition moment is
\begin{equation}
\left|x_{10}^{\mathrm{max}}\right| = \frac{\hbar}{\sqrt{m E_{10}}} \sqrt{\frac{3}{2}\lambda_{00} - 1} .
\label{eq:xmax}
\end{equation}

To find an expression for the off-resonant polarizability, hyperpolarizability, and second hyperpolarizability of a three-level system, we substitute Eqs. \ref{eq:x10fs}-\ref{eq:xmax} into Eqs. \ref{eq:polar}-\ref{eq:sechyper}. The three-level polarizability, hyperpolarizability, and second hyperpolarizability reduce to
\begin{eqnarray}
\alpha_{3L}' &=& \frac{2 e^2 \hbar^2}{m E_{10}^{2}} \left[ X^2 + E^2\left(1 - X^2\right) \right] H_{\lambda}^{2} , \label{eq:alpharel3L}
\end{eqnarray}
\begin{widetext}
\begin{eqnarray}
\beta_{3L}' &=& \frac{6 e^3 \hbar^3}{ \sqrt{m^3 E_{10}^7}} H_\lambda \left[\displaystyle\sqrt{1-X^2} X \left(1 - E\right)^{3/2} \left(1 + \frac{3}{2}E + E^2\right) H_\lambda G_{\lambda}\left(X\right)
- 3 X \lambda_{10} - 3 \displaystyle\sqrt{1 - X^2} E^{7/2} \lambda_{20} \right] ,
\label{eq:betarel3L}
\end{eqnarray}
and
\begin{eqnarray}
\gamma_{3L}' &=& \frac{e^4 \hbar^4}{m^2 E_{10}^{5}} \left\{ 4\left[4 - \left(1 + 2 X^2 + 5 X^4\right) E^5 - \left(1 - 2 X^2 - 5 X^4\right) E^3 - \left(3 - 5 X^4\right) E^2 - 5 X^4\right] \right.\nonumber \\
&-& \left. 9 \left[ \left(1 - 2 X^2 + 5 X^4\right) E^5 + \left(2 X^2 - 5 X^4\right) E^3 + \left(4 X^2 - 5 X^4\right) E^2 + 5X^4 - 4X^2 \right] \lambda_{00}^2 \right. \nonumber \\
&-& \left. 6\left[4 - 4 X^2 + (4 X^2 - 3) E^2 + (4 X^2 - 1) E^3 - 4 X^2 E^5 \right] \lambda_{11} \right. \nonumber \\
&+& \left. 6 \left[ \left(2 + 10 X^4\right) E^5 + \left(1 - 10 X^4\right) E^3 + \left(3 + 4 X^2 - 10 X^4\right) E^2 + 10X^4 - 4X^2 - 4 \right]\lambda_{00} \right. \nonumber \\
&-& \left. 9\left[ 4 X^2 E^5 \left(1- 4X^2\right) E^3 + \left(3 - 4X^2\right) E^2 + 4X^2 - 4\right]\lambda_{00} \lambda_{11} + 9 \left(E^5 \lambda_{20}^2 + \lambda_{10}^2\right) \right. \nonumber \\
&-& \left. 12 H_\lambda G_{\lambda}\left(X\right) \left[ \lambda_{10} \left(2 + E\right)\displaystyle\sqrt{1-E}\displaystyle\sqrt{1-X^2} - \lambda_{20} X \displaystyle\sqrt{E\left(1-E\right)} \left(E^3 + 2 E^4\right) \right] \right\} ,
\label{eq:gammarel3L}
\end{eqnarray}
\hfill{}\hfill{}\hfill{}\hfill{}
\end{widetext}
where
\begin{equation}
H_\lambda = \sqrt{\frac{3}{2}\lambda_{00}-1} .
\label{eq:Hlam}
\end{equation}
The primed coefficients in Eqs. \ref{eq:alpharel3L}-\ref{eq:gammarel3L} denote relativistic corrections to the TRK sum rules. Note that the energies in these primed equations for the nonlinear-optical coefficients are also relativistically corrected.

In the non-relativistic limit, \textit{i}.\textit{e}., when $c \rightarrow \infty$, Eqs. \ref{eq:alpharel3L}-\ref{eq:gammarel3L} reduce to the off-resonant, three-level model calculated from the non-relativistic TRK sum rules.\cite{kuzyk09.01,perez01.08} The polarizability, hyperpolarizability and second hyperpolarizability in the non-relativistic limit are given by
\begin{eqnarray}
\alpha_{3L} &=& \frac{e^2 \hbar^2}{m E_{10}^{2}} \left[X^2 + E^2 \left(1 - X^2 \right) \right] \label{eq:alphanonrel} \\
\beta_{3L} &=& \frac{3 e^3 \hbar^3}{2 \sqrt{2 m^3 E_{10}^{7}}} X \displaystyle\sqrt{1-X^4} \nonumber \\
&\times& \left(1-E\right)^{3/2} \left(1+\frac{3}{2}E+E^2\right)
\label{eq:betanonrel}
\end{eqnarray}
and
\begin{eqnarray}
\gamma_{3L} &=& \frac{e^4 \hbar^4}{m^2 E_{10}^{5}} \left[4 - 2(E^2-1)E^3 X^2 \right. \nonumber \\
&-& 5 \left(E-1\right)^2\left(E+1\right)\left(E^2+E+1\right)X^4 \nonumber \\
&-& \left. \left(E^3+E+3\right)E^2 \right] .
\label{eq:gammanonrel}
\end{eqnarray}

\section{Discussion}

Transition moments (and expectation values of many types) in addition to diagonal energy/potential differences can appear in the relativistically corrected equation via the $\lambda_{ij}$ terms. If the values of $\lambda_{ij}$ are known for a specific potential, then the second hyperpolarizability can be approximated by Eq. \ref{eq:gammarel3L}. In other words, the inclusion of the momentum term in the TRK sum rules no longer gives a simple relationship between the transition moments and energies. 

It is clear that the linear polarizability for all $X$ and $E$ is reduced by the lowest-order relativistic correction. The decrease is due to the presence of the $H_\lambda$ parameter, which can take values between 0 and 1, where $H_\lambda \rightarrow 1$ in the non-relativistic limit. In $X$ and $E$ parameter space, the limit of the hyperpolarizability is located at $X = 1/\sqrt[4]{3}$ and $E = 0$. The resulting limit corresponds to a two-level system, which is not surprising given the relationships in Eq. \ref{eq:00eqfull}. Because that the maximum is located when $1/E_{20} \rightarrow 0$, it seems counterintuitive that the maximum of the nonlinear-optical coefficients occur when $X\neq1$; however, we can no longer think in terms of simple linear optics. When calculating nonlinear-optical coefficients, the intermediate states and excited state sum rules are interwoven into Eqs. \ref{eq:polar} and \ref{eq:sechyper}. The limit of the hyperpolarizability of non-relativistic systems calculated using the three-level ansatz is given by,
\begin{equation}
\beta_{\mathrm{max}} = \sqrt[4]{3}\frac{e^3 \hbar^3}{\sqrt{m^3 E_{10}^{7}}} .
\label{eq:betanonrelupper}
\end{equation}

\begin{figure}[t!]
\centering\includegraphics{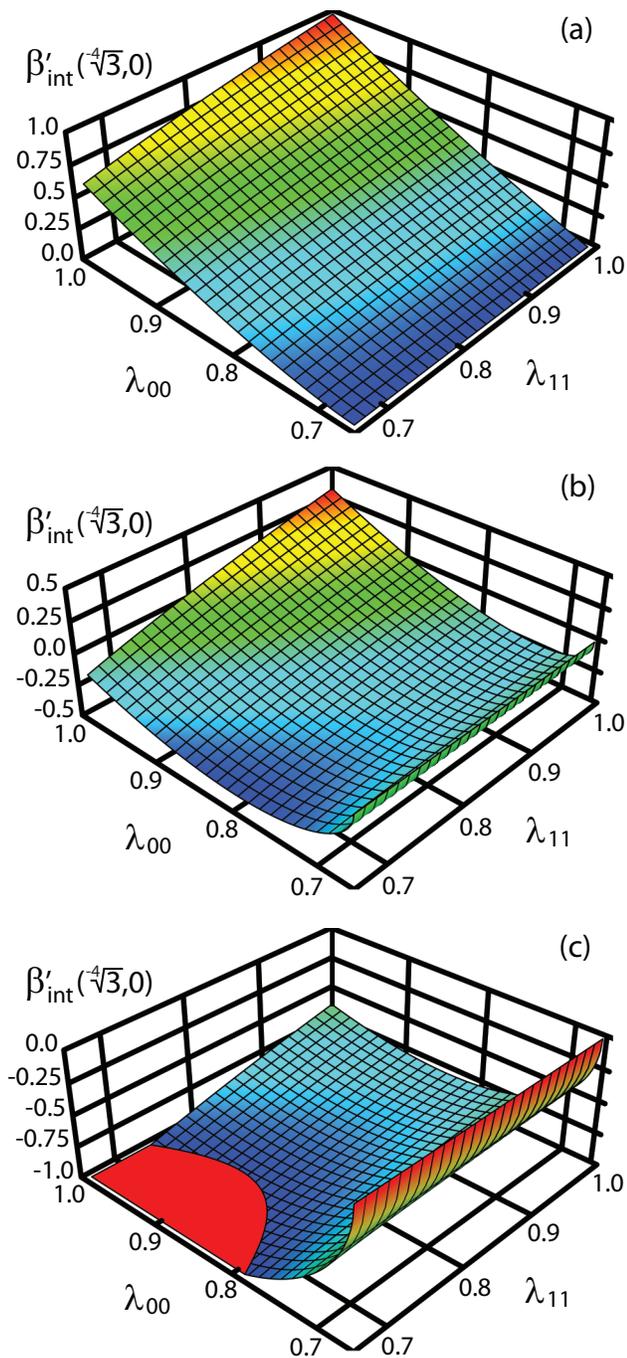}
\caption{The intrinsic hyperpolarizability as a function of $\lambda_{00}$ and $\lambda_{11}$ when $X = 1/\sqrt[4]{3}$ and $E = 0$. The relativistic behavior of the Kuzyk limit is illustrated for (a) $\lambda_{10} = 0$, (b) $\lambda_{10} = 0.1$, and (c) $\lambda_{10} = 0.2$. Values outside the magnitude for the non-relativistic limit are given by the solid (red) region.}
\label{fig:betamax}
\end{figure}

The effects of linear, relativistic, kinetic energy on the fundamental limit of the hyperpolarizability may be studied by substituting $X = 1/\sqrt[4]{3}$ and $E = 0$ into Eq. \ref{eq:betarel3L}. The lowest-order relativistic correction to the limit of the hyperpolarizability, $\beta' \left(X,E\right)$, is given by
\begin{eqnarray}
\beta' \left(\sqrt[-4]{3},0\right) &=& \frac{2}{\sqrt[4]{3}} \frac{e^3 \hbar^3}{\sqrt{m^3 E_{10}^{7}}} \, H_\lambda \label{eq:maxrelsubbeta} \\
&\times& \left( \sqrt{6} H_\lambda \sqrt{\frac{3}{2} \lambda_{00} + \frac{3}{2} \lambda_{11} - 1} - 9\lambda_{10} \right) . \nonumber
\end{eqnarray}
Note that when $E=0$, the second excited state is infinitely large; however, $E_{20}$ does not enter into the oscillator strength corrections as there is no $\lambda_{22}$ term. The same is true for any number of truncated states, where there is no diagonal $\lambda_{pp}$ term for a system truncated to $p$ states. Thus, we may still assume that $E_{20} \rightarrow \infty$ without any obvious negative consequences.

The limit to the hyperpolarizability for increasingly relativistic systems is shown in Fig. \ref{fig:betamax}, where the intrinsic value, $\beta_{int}' = \beta'/\beta_{max}$, is plotted as a function of $\lambda_{00}$ and $\lambda_{11}$. We must place a lower bound on some parameters due to the low-order approximation. We observe that for real values of the off-resonant hyperpolarizability, $\lambda_{00}$ and $\lambda_{11}$ can have a minimum value of $2/3$. As shown in Fig. \ref{fig:betamax}(a), the lowest-order relativistic correction to the limit of the hyperpolarizability is reduced, or even negative, when $\lambda_{10} = 0$ while $\lambda_{00}$ and $\lambda_{11}$ increase. The hyperpolarizability is further reduced when the off-diagonal relativistic term, $\lambda_{10}$, is increased as illustrated in Fig. \ref{fig:betamax}(b). If we further increase $\lambda_{10}$ away from the non-relativistic limit, there are values of $\lambda_{00}$ and $\lambda_{11}$ that correspond to a negative hyperpolarizability that is greater in magnitude than the fundamental limit. These occurrences where the limit is broken appear for values of $\lambda_{11}$ that deviate from unity, but not for large deviations of $\lambda_{00}$, where the entire function of $\beta'$ is multiplied by $H_{\lambda}$. Thus, large values of $\lambda_{00}$ quickly decrease the effects of an increasing $\lambda_{10}$.

The red region shown in Fig. \ref{fig:betamax}(c) corresponds to the region that is opposite in sign and greater in magnitude to the fundamental limit when $\lambda_{10} = 0.2$, which is still within the stability boundaries of the lowest-order approximation. There is the possibility that higher-order relativistic corrections may lessen the effects of the lowest-order correction; however, introducing higher-order corrections into an analytical framework is quite complicated and beyond the scope of the present study. The lowest-order correction to the ($n$,$n$) sum rules appears to damp the total strength of the transition probabilities by increasing the momentum at semi-relativistic energies. Note that an exotic Hamiltonian with a small momentum correction of opposite sign to that of the lowest-order relativistic correction would instead produce a virtual increase in the total oscillator strength. Relativistic corrections to the ($n$,$k$) TRK sum rules, where $n\neq k$, appear to directly subtract from the total response as opposed to an apparent quadratic damping. These nonzero terms are what appear to allow the non-relativistic fundamental limit to be broken when scaled to semi-relativistic kinetic energies.

\begin{figure}[t!]
\centering\includegraphics{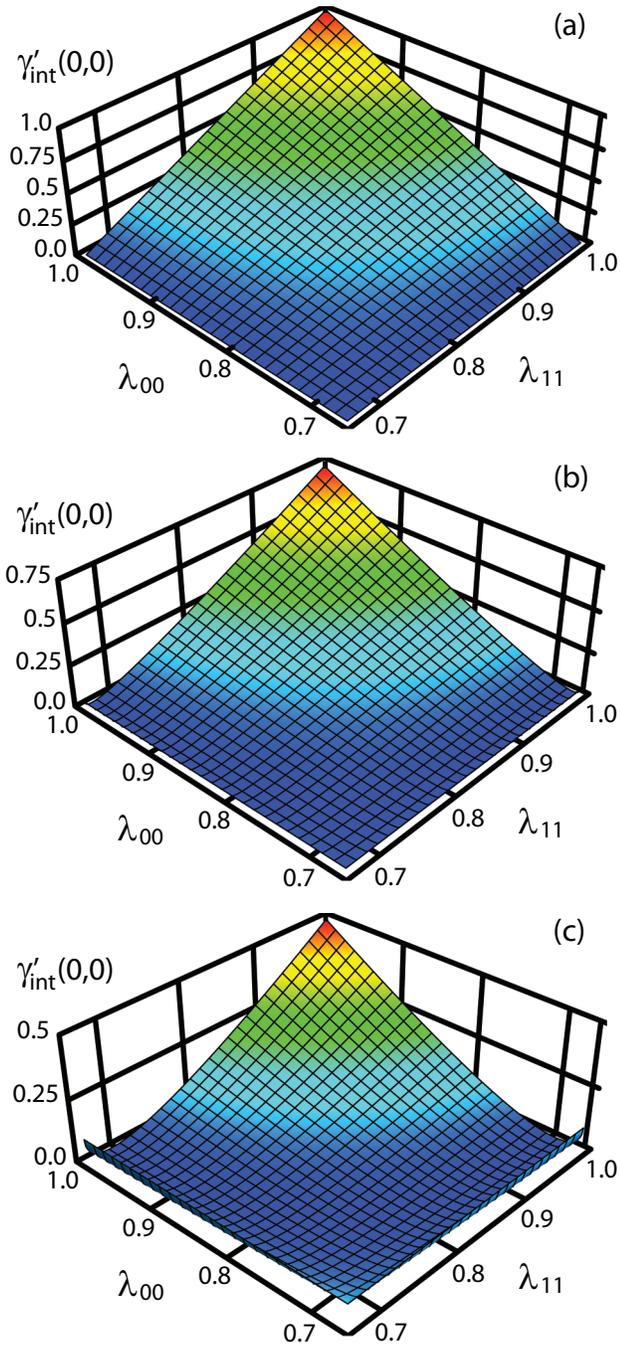}
\caption{The maximum intrinsic value of the second hyperpolarizability as a function of $\lambda_{00}$ and $\lambda_{11}$ from the relativistic corrections to the TRK sum rules. The corrected maximum is shown for (a) $\lambda_{10} = 0$, (b) $\lambda_{10} = 0.1$, and (c) $\lambda_{10} = 0.2$.}
\label{fig:max}
\end{figure}

To get a general idea of how relativity affects the second hyperpolarizability, we first study the limits of the non-relativistic three-level model, Eq. \ref{eq:gammanonrel}. The upper limit of the non-relativistic second hyperpolarizability, in the reduced parameter space, is located at $E=0$ and $X=0$, which gives
\begin{equation}
\gamma_{\mathrm{max}} = \frac{4 e^4 \hbar^4}{m^2 E_{10}^{5}} .
\label{eq:gammanonrelupper}
\end{equation}
The lower limit is found when either $E = 1$, or when $E = 0$ and $X = 1$. For the non-relativistic case, the lower limit of the second hyperpolarizability is
\begin{equation}
\gamma_{\mathrm{min}} = -\frac{e^4 \hbar^4}{m^2 E_{10}^{5}} .
\label{eq:gammanonrellower}
\end{equation}

We may now substitute the corresponding three-level energy and first transition moment fractions, $X$ and $E$, into the lowest-order corrected second hyperpolarizability expression to study the maximum value of semi-relativistic systems. After substituting the parameters associated with the maximum for the non-relativistic limit, Eq. \ref{eq:gammarel3L} reduces to
\begin{eqnarray}
\gamma' \left(0,0\right) &=& \frac{e^4 \hbar^4}{m^2 E_{10}^{5}} \left[16 + 9\lambda_{10} - 24\lambda_{11} + 3 \lambda_{00} \left(12 \lambda_{11} - 8\right) \right. \nonumber \\
&-& \left. 12 \lambda_{10} \sqrt{\left(3\lambda_{00} - 2\right) \left(3\lambda_{11} - 2\right)} \right] .
\label{eq:maxrelsub}
\end{eqnarray}

\begin{figure}[t!]
\centering\includegraphics{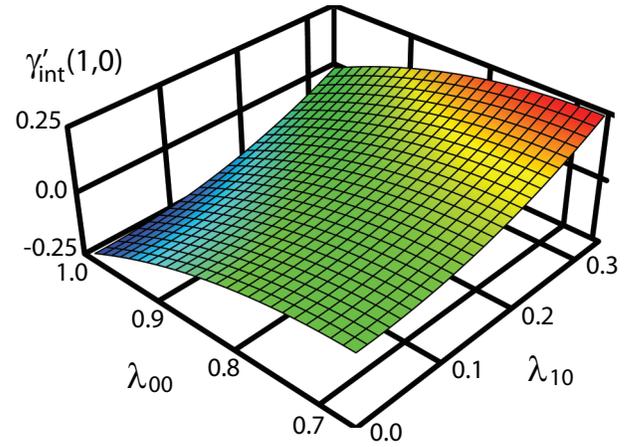}
\caption{The minimum intrinsic value of the second hyperpolarizability as a function of $\lambda_{00}$ and $\lambda_{10}$ from the relativistic corrections to the TRK sum rules.}
\label{fig:min}
\end{figure}

The lowest-order relativistic corrections to the second hyperpolarizability are shown in Fig. \ref{fig:max}. Note that the maximum intrinsic value, $\gamma_{int}'$, is $1$ and the minimum is $-1/4$. 
As shown in Fig. \ref{fig:max}(a), the maximum possible second hyperpolarizability is reduced for a potential with negligible off-diagonal $\lambda$ parameters. The other two plots in Fig. \ref{fig:max} illustrate how a nonzero $\lambda_{10}$ further reduces the second nonlinear response from the non-relativistic maximum. Again, note that even though the intrinsic values are reduced, the net numerical values for the off-resonant response may be affected differently because of relativistic changes in $E_{10}$.

There are two regimes that are associated with the minimum value of the second hyperpolarizability. Focusing only on the minimum at the two-level limit, \textit{i}.\textit{e}. $E \rightarrow 0$, there is an intrinsic value of $-0.25$ when $X = 1$. The lower limit in this regime, with lowest-order relativistic corrections, is given by
\begin{equation}
\gamma' \left(1,0\right) = \frac{e^4 \hbar^4}{m^2 E_{10}^{5}} \left[12 \lambda_{00} - 9 \lambda_{00}^2 + 9 \lambda_{10}^2 - 4 \right] .
\label{eq:minrelsub}
\end{equation}
The minimum value in this regime is only affected by the lowest diagonal term, $\lambda_{00}$ and the first off-diagonal term, $\lambda_{10}$. Thus, it appears that, for well-behaved systems under these approximations, the first excited state does not contribute to the lowest-order relativistic correction at the (1,0) minimum.

The minimum at (1,0) is plotted in Fig. \ref{fig:min} as a function of $\lambda_{00}$ and $\lambda_{10}$. The value of $\lambda_{00}$ is `walked' away from the non-relativistic value of 1, while $\lambda_{10}$ is increased from the non-relativistic limit of zero. Notice how the magnitude of the lower limit in this regime is also reduced which signifies response damping as the dominant mechanism as opposed to a subtraction of the net response. Similar to $\beta'$, under more extreme circumstances, it appears that a negative value of $\gamma$ may also become zero or even positive. The positive is due to the off-diagonal $\lambda_{kn}$ subtraction of the response which is less prominent for the second hyperpolarizability. 

The second regime where there exists a minimum is found when $E = 1$, where the minimum also reaches the negative intrinsic limit of $-1/4$. In this regime, the lowest-order relativistic correction gives
\begin{equation}
\gamma' \left(X,1\right) = \frac{e^4 \hbar^4}{m^2 E_{10}^{5}} \left[12 \lambda_{00} - 9 \lambda_{00}^2 + 9 \lambda_{10}^2 + 9 \lambda_{20}^2 - 4 \right] .
\label{eq:minDrelsub}
\end{equation}
Thus, this degenerate minimum is more strongly affected by relativistic corrections with the inclusion of a positive $\lambda_{20}$ parameter that increases the minimum value away from the negative limit. 

\vspace{0.1cm}

\begin{center}\textit{Relativistic effects of H-like ions and the 3-level ansatz}\end{center}

\vspace{0.1cm}


It is well-known that, unlike many organic molecules, the continuum states make a significant contribution to the total dipole response of a single hydrogen atom. Thus, these continuum states cause problems with the SOS method for the second hyperpolarizability. Other non-relativistic methods have been developed such as a time-independent perturbation approach \cite{sewel49.01,boyle01.66} to calculate the zero-frequency response and a method employing Sturmian functions used by Shelton \cite{shelt87.03} to calculate the frequency-dependent coefficients. Because the largest portion of the dipole strength is in the $1$s-$2$p transition, a qualitatively study of the lowest-relativistic corrections to H-like ions can be performed with a simple three-level model. Here, problems with convergence and continuum states are washed away by placing the entire oscillator strength in the first two excited state transitions, which gives an reasonably approximate description for most systems.

The only nonzero angular contributions from the non-relativistic transitions are either $1/\sqrt{3}$ for $n$s-$n'$p or $2/\sqrt{15}$ for $n$p-$n'$d. The similarity between the two nonzero angular contributions, the low frequency of $n$p-$n'$d transitions in the SOS expression, and the fact that we can limit our study to the $\gamma_{zzzz}$ component allows us to make a one-dimensional approximation; thus, the total response is given by Eq. \ref{eq:1DrelTRK}. Note that the spin-orbit term will still enter into the calculation, but it will be later introduced as a perturbation in the energy so that we can further simplify the example.

We can treat the lowest-order linear moment, spin-orbit, and Darwin terms as first order perturbations in the energy.\cite{griff95.01} This provides a simpler approach when solving Eq. \ref{eq:simpleham}, where we may write
\begin{eqnarray}
\left(p_{nk}^{\mathrm{H-like}}\right)^2 &\approx& \frac{Z^4 \alpha^2}{\left(n+1\right)^2} \left[\frac{2}{\left(1+2j\right)\left(n+1\right)} \right. \nonumber \\
&-& \left. \frac{3}{4 \left(n+1\right)^2} \right] - \frac{Z^2}{\left(n+1\right)^2} - 2 V_{nk}
\label{eq:pnkHapprox}
\end{eqnarray}
given in atomic units ($\hbar\rightarrow 1$, $m\rightarrow 1$, $e\rightarrow 1$), where $Z$ is the number of protons, $\alpha$ is the fine structure constant, $n=0,1,2,\cdots$, and $V = -Z/r$. We can simplify the example even further by assuming a single energy level from averaging the $j=l\pm 1/2$ splitting for $l\neq 0$. To make this simplification, the transition probabilities for the $1$s$_{1/2}$-$1$p$_{1/2}$ and $1$s$_{1/2}$-$1$p$_{3/2}$ doublet as well as for the second transition's $2$p$_{1/2}$-$3$p$_{3/2}$, $2$p$_{3/2}$-$3$p$_{3/2}$, and $2$p$_{3/2}$-$3$p$_{5/2}$ multiplet are evaluated from the Dirac equation,\cite{Bethe77.01,garst01.71,young75.01} and used to perform the weighted averages for the excited state energies.

The $\lambda_{kn}$ terms are then given by
\begin{equation}
\lambda_{nk}^{\mathrm{H-like}} \approx \delta_{nk} - \alpha^2 \left(p_{nk}^{\mathrm{H-like}}\right)^2 ,
\label{eq:lambdaHapprox}
\end{equation}
where the off-diagonal terms in Eq. \ref{eq:lambdaHapprox} are taken to be zero under the current set of approximations with energy perturbations from a three-dimensional central potential.

\begin{figure}[t!]
\centering\includegraphics{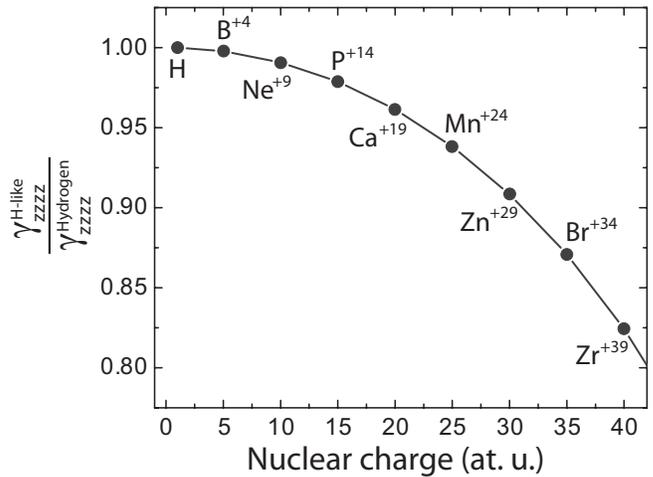}
\caption{The second hyperpolarizability as a function of atomic number for an H-like ion using the three-level model with lowest-order relativistic corrections to the TRK sum rules.}
\label{fig:gammaapprox}
\end{figure}

The second hyperpolarizability can now be calculated by substituting the approximate Z-dependent coefficients $E_{10}^{\mathrm{H-like}}$, $E^{\mathrm{H-like}}$, $X^{\mathrm{H-like}}$, $\lambda_{01}^{\mathrm{H-like}}$, and $\lambda_{02}^{\mathrm{H-like}}$, into Eq. \ref{eq:gammarel3L}. The ratio of the second hyperpolarizability for H-like atoms for the $z$-diagonal tensor component, $\lambda^{\mathrm{H-like}}_{zzzz}$, divided by the approximate second hyperpolarizability of the hydrogen atom, $\lambda^{\mathrm{hydrogen}}_{zzzz}$, is given in Fig. \ref{fig:gammaapprox} as a function of the atomic number. The total strength of the transitions decrease causing a drop in the static nonlinear optical response. 
Note that the severity of damping to the second nonlinear response is lessened by a decrease in the first transition energy as the atomic number increases.

\section{Conclusion}

The lowest-order relativistic correction to the TRK sum rules was shown to limit the oscillator strength below the value derived from the non-relativistic Hamiltonian. This correction was applied to both the static linear and first two nonlinear optical responses; the magnitude of this correction is no longer a constant and depends on the potential energy function. This lowest-order relativistic correction has been applied to the three-level ansatz, where in the relativistic regime, the magnitudes of the fundamental limits of the polarizability, hyperpolarizability, and second hyperpolarizability are reduced. Thus, the non-relativistic regime gives the largest values of the fundamental limit for closed quantum systems.

In the regime where the relativistic parameters pull the hyperpolarizability, at the positive fundamental limit, to below the negative bound, we find that it may be possible to break the Kuzyk limit (although with opposite sign). This is a disturbing result and further studies with higher degrees of accuracy must be performed for this consequence to be supported. Further studies with additional corrections may also help in understanding the peculiar influences of the off-diagonal sum rules on the linear and nonlinear responses. Originally, these off-diagonal terms were equal to zero in the non-relativistic limit, where they are the primary reason that the Kuzyk limit of the hyperpolarizability could possibly be broken when referencing to the lowest-order correction.

\vspace{0.05cm}

\noindent {\bf Acknowledgments}

\vspace{0.05cm}

\acknowledgments I would like to thank Prof. Kenneth D. Singer and Prof. Mark G. Kuzyk for useful discussions. I would also like to thank the National Science Foundation grant number OISE-1243313 for partial support of this project.

\hfill{}


\begin{thebibliography}{64}%
\makeatletter
\providecommand \@ifxundefined [1]{%
 \@ifx{#1\undefined}
}%
\providecommand \@ifnum [1]{%
 \ifnum #1\expandafter \@firstoftwo
 \else \expandafter \@secondoftwo
 \fi
}%
\providecommand \@ifx [1]{%
 \ifx #1\expandafter \@firstoftwo
 \else \expandafter \@secondoftwo
 \fi
}%
\providecommand \natexlab [1]{#1}%
\providecommand \enquote  [1]{``#1''}%
\providecommand \bibnamefont  [1]{#1}%
\providecommand \bibfnamefont [1]{#1}%
\providecommand \citenamefont [1]{#1}%
\providecommand \href@noop [0]{\@secondoftwo}%
\providecommand \href [0]{\begingroup \@sanitize@url \@href}%
\providecommand \@href[1]{\@@startlink{#1}\@@href}%
\providecommand \@@href[1]{\endgroup#1\@@endlink}%
\providecommand \@sanitize@url [0]{\catcode `\\12\catcode `\$12\catcode
  `\&12\catcode `\#12\catcode `\^12\catcode `\_12\catcode `\%12\relax}%
\providecommand \@@startlink[1]{}%
\providecommand \@@endlink[0]{}%
\providecommand \url  [0]{\begingroup\@sanitize@url \@url }%
\providecommand \@url [1]{\endgroup\@href {#1}{\urlprefix }}%
\providecommand \urlprefix  [0]{URL }%
\providecommand \Eprint [0]{\href }%
\providecommand \doibase [0]{http://dx.doi.org/}%
\providecommand \selectlanguage [0]{\@gobble}%
\providecommand \bibinfo  [0]{\@secondoftwo}%
\providecommand \bibfield  [0]{\@secondoftwo}%
\providecommand \translation [1]{[#1]}%
\providecommand \BibitemOpen [0]{}%
\providecommand \bibitemStop [0]{}%
\providecommand \bibitemNoStop [0]{.\EOS\space}%
\providecommand \EOS [0]{\spacefactor3000\relax}%
\providecommand \BibitemShut  [1]{\csname bibitem#1\endcsname}%
\let\auto@bib@innerbib\@empty
\bibitem [{\citenamefont {Kuzyk}(2000{\natexlab{a}})}]{kuzyk00.01}%
  \BibitemOpen
  \bibfield  {author} {\bibinfo {author} {\bibfnamefont {M.~G.}\ \bibnamefont
  {Kuzyk}},\ }\href@noop {} {\bibfield  {journal} {\bibinfo  {journal} {Phys.
  Rev. Lett.}\ }\textbf {\bibinfo {volume} {85}},\ \bibinfo {pages} {1218}
  (\bibinfo {year} {2000}{\natexlab{a}})}\BibitemShut {NoStop}%
\bibitem [{\citenamefont {Thomas}(1925)}]{thoma25.01}%
  \BibitemOpen
  \bibfield  {author} {\bibinfo {author} {\bibfnamefont {W.}~\bibnamefont
  {Thomas}},\ }\href@noop {} {\bibfield  {journal} {\bibinfo  {journal}
  {Naturwissenschaften}\ }\textbf {\bibinfo {volume} {13}},\ \bibinfo {pages}
  {627} (\bibinfo {year} {1925})}\BibitemShut {NoStop}%
\bibitem [{\citenamefont {Reiche}\ and\ \citenamefont
  {Thomas}(1925)}]{reich25.01}%
  \BibitemOpen
  \bibfield  {author} {\bibinfo {author} {\bibfnamefont {F.}~\bibnamefont
  {Reiche}}\ and\ \bibinfo {author} {\bibfnamefont {W.}~\bibnamefont
  {Thomas}},\ }\href@noop {} {\bibfield  {journal} {\bibinfo  {journal} {Z.
  Phys.}\ }\textbf {\bibinfo {volume} {34}},\ \bibinfo {pages} {510} (\bibinfo
  {year} {1925})}\BibitemShut {NoStop}%
\bibitem [{\citenamefont {Kuhn}(1925)}]{kuhn25.01}%
  \BibitemOpen
  \bibfield  {author} {\bibinfo {author} {\bibfnamefont {W.}~\bibnamefont
  {Kuhn}},\ }\href@noop {} {\bibfield  {journal} {\bibinfo  {journal} {Z.
  Phys.}\ }\textbf {\bibinfo {volume} {33}},\ \bibinfo {pages} {408} (\bibinfo
  {year} {1925})}\BibitemShut {NoStop}%
\bibitem [{\citenamefont {Kuzyk}(2000{\natexlab{b}})}]{kuzyk00.02}%
  \BibitemOpen
  \bibfield  {author} {\bibinfo {author} {\bibfnamefont {M.~G.}\ \bibnamefont
  {Kuzyk}},\ }\href@noop {} {\bibfield  {journal} {\bibinfo  {journal} {Opt.
  Lett.}\ }\textbf {\bibinfo {volume} {25}},\ \bibinfo {pages} {1183} (\bibinfo
  {year} {2000}{\natexlab{b}})}\BibitemShut {NoStop}%
\bibitem [{\citenamefont {Kuzyk}(2001)}]{kuzyk01.01}%
  \BibitemOpen
  \bibfield  {author} {\bibinfo {author} {\bibfnamefont {M.~G.}\ \bibnamefont
  {Kuzyk}},\ }\href@noop {} {\bibfield  {journal} {\bibinfo  {journal} {IEEE
  Journal on Selected Topics in Quantum Electronics}\ }\textbf {\bibinfo
  {volume} {7}},\ \bibinfo {pages} {774 } (\bibinfo {year} {2001})}\BibitemShut
  {NoStop}%
\bibitem [{\citenamefont {Clays}(2001)}]{clays01.01}%
  \BibitemOpen
  \bibfield  {author} {\bibinfo {author} {\bibfnamefont {K.}~\bibnamefont
  {Clays}},\ }\href@noop {} {\bibfield  {journal} {\bibinfo  {journal} {Opt.
  Lett.}\ }\textbf {\bibinfo {volume} {26}},\ \bibinfo {pages} {1699} (\bibinfo
  {year} {2001})}\BibitemShut {NoStop}%
\bibitem [{\citenamefont {Kuzyk}(2003)}]{kuzyk03.03}%
  \BibitemOpen
  \bibfield  {author} {\bibinfo {author} {\bibfnamefont {M.~G.}\ \bibnamefont
  {Kuzyk}},\ }\href@noop {} {\bibfield  {journal} {\bibinfo  {journal} {J. Chem
  Phys.}\ }\textbf {\bibinfo {volume} {119}},\ \bibinfo {pages} {8327}
  (\bibinfo {year} {2003})}\BibitemShut {NoStop}%
\bibitem [{\citenamefont {Clays}\ and\ \citenamefont {Coe}(2003)}]{clays03.01}%
  \BibitemOpen
  \bibfield  {author} {\bibinfo {author} {\bibfnamefont {K.}~\bibnamefont
  {Clays}}\ and\ \bibinfo {author} {\bibfnamefont {B.~J.}\ \bibnamefont
  {Coe}},\ }\href@noop {} {\bibfield  {journal} {\bibinfo  {journal} {Chem.
  Mater.}\ }\textbf {\bibinfo {volume} {15}},\ \bibinfo {pages} {642} (\bibinfo
  {year} {2003})}\BibitemShut {NoStop}%
\bibitem [{\citenamefont {Tripathi}\ \emph {et~al.}(2004)\citenamefont
  {Tripathi}, \citenamefont {Moreno}, \citenamefont {Kuzyk}, \citenamefont
  {Coe}, \citenamefont {Clays},\ and\ \citenamefont {Kelley}}]{Tripa04.01}%
  \BibitemOpen
  \bibfield  {author} {\bibinfo {author} {\bibfnamefont {K.}~\bibnamefont
  {Tripathi}}, \bibinfo {author} {\bibfnamefont {P.}~\bibnamefont {Moreno}},
  \bibinfo {author} {\bibfnamefont {M.~G.}\ \bibnamefont {Kuzyk}}, \bibinfo
  {author} {\bibfnamefont {B.~J.}\ \bibnamefont {Coe}}, \bibinfo {author}
  {\bibfnamefont {K.}~\bibnamefont {Clays}}, \ and\ \bibinfo {author}
  {\bibfnamefont {A.~M.}\ \bibnamefont {Kelley}},\ }\href@noop {} {\bibfield
  {journal} {\bibinfo  {journal} {J. Chem. Phys.}\ }\textbf {\bibinfo {volume}
  {121}},\ \bibinfo {pages} {7932} (\bibinfo {year} {2004})}\BibitemShut
  {NoStop}%
\bibitem [{\citenamefont {P\`{e}rez~Moreno}\ and\ \citenamefont
  {Kuzyk}(2005)}]{perez05.01}%
  \BibitemOpen
  \bibfield  {author} {\bibinfo {author} {\bibfnamefont {J.}~\bibnamefont
  {P\`{e}rez~Moreno}}\ and\ \bibinfo {author} {\bibfnamefont {M.~G.}\
  \bibnamefont {Kuzyk}},\ }\href@noop {} {\bibfield  {journal} {\bibinfo
  {journal} {J. Chem. Phys.}\ }\textbf {\bibinfo {volume} {123}},\ \bibinfo
  {pages} {194101} (\bibinfo {year} {2005})}\BibitemShut {NoStop}%
\bibitem [{\citenamefont {Kuzyk}(2005)}]{kuzyk05.02}%
  \BibitemOpen
  \bibfield  {author} {\bibinfo {author} {\bibfnamefont {M.~G.}\ \bibnamefont
  {Kuzyk}},\ }\href@noop {} {\bibfield  {journal} {\bibinfo  {journal} {Phys.
  Rev. A.}\ }\textbf {\bibinfo {volume} {72}},\ \bibinfo {pages} {053819}
  (\bibinfo {year} {2005})}\BibitemShut {NoStop}%
\bibitem [{\citenamefont {Kuzyk}(2006{\natexlab{a}})}]{kuzyk06.01}%
  \BibitemOpen
  \bibfield  {author} {\bibinfo {author} {\bibfnamefont {M.~G.}\ \bibnamefont
  {Kuzyk}},\ }\href@noop {} {\bibfield  {journal} {\bibinfo  {journal} {J.
  Nonl. Opt. Phys. \& Mat.}\ }\textbf {\bibinfo {volume} {15}},\ \bibinfo
  {pages} {77} (\bibinfo {year} {2006}{\natexlab{a}})}\BibitemShut {NoStop}%
\bibitem [{\citenamefont {Kuzyk}(2006{\natexlab{b}})}]{kuzyk06.03}%
  \BibitemOpen
  \bibfield  {author} {\bibinfo {author} {\bibfnamefont {M.~G.}\ \bibnamefont
  {Kuzyk}},\ }\href@noop {} {\bibfield  {journal} {\bibinfo  {journal} {J. Chem
  Phys.}\ }\textbf {\bibinfo {volume} {125}},\ \bibinfo {pages} {154108}
  (\bibinfo {year} {2006}{\natexlab{b}})}\BibitemShut {NoStop}%
\bibitem [{\citenamefont {P\'{e}rez~Moreno}\ \emph
  {et~al.}(2007{\natexlab{a}})\citenamefont {P\'{e}rez~Moreno}, \citenamefont
  {Zhao}, \citenamefont {Clays},\ and\ \citenamefont {Kuzyk}}]{perez07.01}%
  \BibitemOpen
  \bibfield  {author} {\bibinfo {author} {\bibfnamefont {J.}~\bibnamefont
  {P\'{e}rez~Moreno}}, \bibinfo {author} {\bibfnamefont {Y.}~\bibnamefont
  {Zhao}}, \bibinfo {author} {\bibfnamefont {K.}~\bibnamefont {Clays}}, \ and\
  \bibinfo {author} {\bibfnamefont {M.~G.}\ \bibnamefont {Kuzyk}},\ }\href@noop
  {} {\bibfield  {journal} {\bibinfo  {journal} {Opt. Lett.}\ }\textbf
  {\bibinfo {volume} {32}},\ \bibinfo {pages} {59} (\bibinfo {year}
  {2007}{\natexlab{a}})}\BibitemShut {NoStop}%
\bibitem [{\citenamefont {P\'{e}rez~Moreno}\ \emph
  {et~al.}(2007{\natexlab{b}})\citenamefont {P\'{e}rez~Moreno}, \citenamefont
  {Asselberghs}, \citenamefont {Zhao}, \citenamefont {Song}, \citenamefont
  {Nakanishi}, \citenamefont {Okada}, \citenamefont {Nogi}, \citenamefont
  {Kim}, \citenamefont {Je}, \citenamefont {Matrai}, \citenamefont {De~Mayer},\
  and\ \citenamefont {Kuzyk}}]{perez07.02}%
  \BibitemOpen
  \bibfield  {author} {\bibinfo {author} {\bibfnamefont {J.}~\bibnamefont
  {P\'{e}rez~Moreno}}, \bibinfo {author} {\bibfnamefont {I.}~\bibnamefont
  {Asselberghs}}, \bibinfo {author} {\bibfnamefont {Y.}~\bibnamefont {Zhao}},
  \bibinfo {author} {\bibfnamefont {K.}~\bibnamefont {Song}}, \bibinfo {author}
  {\bibfnamefont {H.}~\bibnamefont {Nakanishi}}, \bibinfo {author}
  {\bibfnamefont {S.}~\bibnamefont {Okada}}, \bibinfo {author} {\bibfnamefont
  {K.}~\bibnamefont {Nogi}}, \bibinfo {author} {\bibfnamefont {O.-K.}\
  \bibnamefont {Kim}}, \bibinfo {author} {\bibfnamefont {J.}~\bibnamefont
  {Je}}, \bibinfo {author} {\bibfnamefont {J.}~\bibnamefont {Matrai}}, \bibinfo
  {author} {\bibfnamefont {M.}~\bibnamefont {De~Mayer}}, \ and\ \bibinfo
  {author} {\bibfnamefont {M.~G.}\ \bibnamefont {Kuzyk}},\ }\href@noop {}
  {\bibfield  {journal} {\bibinfo  {journal} {J. Chem. Phys.}\ }\textbf
  {\bibinfo {volume} {126}},\ \bibinfo {pages} {074705} (\bibinfo {year}
  {2007}{\natexlab{b}})}\BibitemShut {NoStop}%
\bibitem [{\citenamefont {P\`{e}rez~Moreno}\ \emph {et~al.}(2008)\citenamefont
  {P\`{e}rez~Moreno}, \citenamefont {Clays},\ and\ \citenamefont
  {Kuzyk}}]{perez01.08}%
  \BibitemOpen
  \bibfield  {author} {\bibinfo {author} {\bibfnamefont {X.}~\bibnamefont
  {P\`{e}rez~Moreno}}, \bibinfo {author} {\bibfnamefont {K.}~\bibnamefont
  {Clays}}, \ and\ \bibinfo {author} {\bibfnamefont {M.~G.}\ \bibnamefont
  {Kuzyk}},\ }\href@noop {} {\bibfield  {journal} {\bibinfo  {journal} {J.
  Chem. Phys.}\ }\textbf {\bibinfo {volume} {128}},\ \bibinfo {pages} {084109}
  (\bibinfo {year} {2008})}\BibitemShut {NoStop}%
\bibitem [{\citenamefont {Zhou}\ and\ \citenamefont {Kuzyk}(2008)}]{zhou08.01}%
  \BibitemOpen
  \bibfield  {author} {\bibinfo {author} {\bibfnamefont {J.}~\bibnamefont
  {Zhou}}\ and\ \bibinfo {author} {\bibfnamefont {M.~G.}\ \bibnamefont
  {Kuzyk}},\ }\href@noop {} {\bibfield  {journal} {\bibinfo  {journal} {J.
  Phys. Chem. C.}\ }\textbf {\bibinfo {volume} {112}},\ \bibinfo {pages} {7978}
  (\bibinfo {year} {2008})}\BibitemShut {NoStop}%
\bibitem [{\citenamefont {Kuzyk}(2009)}]{kuzyk09.01}%
  \BibitemOpen
  \bibfield  {author} {\bibinfo {author} {\bibfnamefont {M.~G.}\ \bibnamefont
  {Kuzyk}},\ }\href@noop {} {\bibfield  {journal} {\bibinfo  {journal} {J.
  Mater. Chem.}\ }\textbf {\bibinfo {volume} {19}},\ \bibinfo {pages}
  {7444–7465} (\bibinfo {year} {2009})}\BibitemShut {NoStop}%
\bibitem [{\citenamefont {Dawson}\ \emph
  {et~al.}(2011{\natexlab{a}})\citenamefont {Dawson}, \citenamefont {Anderson},
  \citenamefont {Schei},\ and\ \citenamefont {Kuzyk}}]{dawson11.02}%
  \BibitemOpen
  \bibfield  {author} {\bibinfo {author} {\bibfnamefont {N.~J.}\ \bibnamefont
  {Dawson}}, \bibinfo {author} {\bibfnamefont {B.~R.}\ \bibnamefont
  {Anderson}}, \bibinfo {author} {\bibfnamefont {J.~L.}\ \bibnamefont {Schei}},
  \ and\ \bibinfo {author} {\bibfnamefont {M.~G.}\ \bibnamefont {Kuzyk}},\
  }\href@noop {} {\bibfield  {journal} {\bibinfo  {journal} {Phys. Rev. A}\
  }\textbf {\bibinfo {volume} {84}},\ \bibinfo {pages} {043406} (\bibinfo
  {year} {2011}{\natexlab{a}})}\BibitemShut {NoStop}%
\bibitem [{\citenamefont {Dawson}\ \emph
  {et~al.}(2011{\natexlab{b}})\citenamefont {Dawson}, \citenamefont {Anderson},
  \citenamefont {Schei},\ and\ \citenamefont {Kuzyk}}]{dawson11.03}%
  \BibitemOpen
  \bibfield  {author} {\bibinfo {author} {\bibfnamefont {N.~J.}\ \bibnamefont
  {Dawson}}, \bibinfo {author} {\bibfnamefont {B.~R.}\ \bibnamefont
  {Anderson}}, \bibinfo {author} {\bibfnamefont {J.~L.}\ \bibnamefont {Schei}},
  \ and\ \bibinfo {author} {\bibfnamefont {M.~G.}\ \bibnamefont {Kuzyk}},\
  }\href@noop {} {\bibfield  {journal} {\bibinfo  {journal} {Phys Rev. A}\
  }\textbf {\bibinfo {volume} {84}},\ \bibinfo {pages} {043407} (\bibinfo
  {year} {2011}{\natexlab{b}})}\BibitemShut {NoStop}%
\bibitem [{\citenamefont {Shafei}\ and\ \citenamefont
  {Kuzyk}(2013)}]{shafe13.01}%
  \BibitemOpen
  \bibfield  {author} {\bibinfo {author} {\bibfnamefont {S.}~\bibnamefont
  {Shafei}}\ and\ \bibinfo {author} {\bibfnamefont {M.~G.}\ \bibnamefont
  {Kuzyk}},\ }\href@noop {} {\bibfield  {journal} {\bibinfo  {journal} {Phys.
  Rev. A}\ }\textbf {\bibinfo {volume} {88}},\ \bibinfo {pages} {023863}
  (\bibinfo {year} {2013})}\BibitemShut {NoStop}%
\bibitem [{\citenamefont {Kuzyk}\ \emph {et~al.}(2013)\citenamefont {Kuzyk},
  \citenamefont {P\'{e}rez-Moreno},\ and\ \citenamefont {Shafei}}]{kuzyk13.01}%
  \BibitemOpen
  \bibfield  {author} {\bibinfo {author} {\bibfnamefont {M.~G.}\ \bibnamefont
  {Kuzyk}}, \bibinfo {author} {\bibfnamefont {J.}~\bibnamefont
  {P\'{e}rez-Moreno}}, \ and\ \bibinfo {author} {\bibfnamefont
  {S.}~\bibnamefont {Shafei}},\ }\href@noop {} {\bibfield  {journal} {\bibinfo
  {journal} {Phys. Rep.}\ }\textbf {\bibinfo {volume} {529}},\ \bibinfo {pages}
  {297} (\bibinfo {year} {2013})}\BibitemShut {NoStop}%
\bibitem [{\citenamefont {Zhou}\ \emph {et~al.}(2006)\citenamefont {Zhou},
  \citenamefont {Kuzyk},\ and\ \citenamefont {Watkins}}]{zhou06.01}%
  \BibitemOpen
  \bibfield  {author} {\bibinfo {author} {\bibfnamefont {J.}~\bibnamefont
  {Zhou}}, \bibinfo {author} {\bibfnamefont {M.~G.}\ \bibnamefont {Kuzyk}}, \
  and\ \bibinfo {author} {\bibfnamefont {D.~S.}\ \bibnamefont {Watkins}},\
  }\href@noop {} {\bibfield  {journal} {\bibinfo  {journal} {Opt. Lett.}\
  }\textbf {\bibinfo {volume} {31}},\ \bibinfo {pages} {2891} (\bibinfo {year}
  {2006})}\BibitemShut {NoStop}%
\bibitem [{\citenamefont {Zhou}\ \emph {et~al.}(2007)\citenamefont {Zhou},
  \citenamefont {Szafruga}, \citenamefont {Watkins},\ and\ \citenamefont
  {Kuzyk}}]{zhou07.02}%
  \BibitemOpen
  \bibfield  {author} {\bibinfo {author} {\bibfnamefont {J.}~\bibnamefont
  {Zhou}}, \bibinfo {author} {\bibfnamefont {U.~B.}\ \bibnamefont {Szafruga}},
  \bibinfo {author} {\bibfnamefont {D.~S.}\ \bibnamefont {Watkins}}, \ and\
  \bibinfo {author} {\bibfnamefont {M.~G.}\ \bibnamefont {Kuzyk}},\ }\href@noop
  {} {\bibfield  {journal} {\bibinfo  {journal} {Phys. Rev. A}\ }\textbf
  {\bibinfo {volume} {76}},\ \bibinfo {pages} {053831} (\bibinfo {year}
  {2007})}\BibitemShut {NoStop}%
\bibitem [{\citenamefont {Wiggers}\ and\ \citenamefont
  {Petschek}(2007)}]{wigge07.01}%
  \BibitemOpen
  \bibfield  {author} {\bibinfo {author} {\bibfnamefont {G.~A.}\ \bibnamefont
  {Wiggers}}\ and\ \bibinfo {author} {\bibfnamefont {R.~G.}\ \bibnamefont
  {Petschek}},\ }\href@noop {} {\bibfield  {journal} {\bibinfo  {journal} {Opt.
  Lett.}\ }\textbf {\bibinfo {volume} {32}},\ \bibinfo {pages} {942} (\bibinfo
  {year} {2007})}\BibitemShut {NoStop}%
\bibitem [{\citenamefont {Kuzyk}\ and\ \citenamefont
  {Kuzyk}(2008)}]{kuzyk08.01}%
  \BibitemOpen
  \bibfield  {author} {\bibinfo {author} {\bibfnamefont {M.~C.}\ \bibnamefont
  {Kuzyk}}\ and\ \bibinfo {author} {\bibfnamefont {M.~G.}\ \bibnamefont
  {Kuzyk}},\ }\href@noop {} {\bibfield  {journal} {\bibinfo  {journal} {J. Opt.
  Soc. Am. B.}\ }\textbf {\bibinfo {volume} {25}},\ \bibinfo {pages} {103}
  (\bibinfo {year} {2008})}\BibitemShut {NoStop}%
\bibitem [{\citenamefont {Watkins}\ and\ \citenamefont
  {Kuzyk}(2009)}]{watkins09.01}%
  \BibitemOpen
  \bibfield  {author} {\bibinfo {author} {\bibfnamefont {D.~S.}\ \bibnamefont
  {Watkins}}\ and\ \bibinfo {author} {\bibfnamefont {M.~G.}\ \bibnamefont
  {Kuzyk}},\ }\href@noop {} {\bibfield  {journal} {\bibinfo  {journal} {J.
  Chem. Phys.}\ }\textbf {\bibinfo {volume} {131}},\ \bibinfo {pages} {064110}
  (\bibinfo {year} {2009})}\BibitemShut {NoStop}%
\bibitem [{\citenamefont {Shafei}\ \emph {et~al.}(2010)\citenamefont {Shafei},
  \citenamefont {Kuzyk},\ and\ \citenamefont {Kuzky}}]{shafe10.01}%
  \BibitemOpen
  \bibfield  {author} {\bibinfo {author} {\bibfnamefont {S.}~\bibnamefont
  {Shafei}}, \bibinfo {author} {\bibfnamefont {M.~C.}\ \bibnamefont {Kuzyk}}, \
  and\ \bibinfo {author} {\bibfnamefont {M.~G.}\ \bibnamefont {Kuzky}},\
  }\href@noop {} {\bibfield  {journal} {\bibinfo  {journal} {J. Opt. Soc. Am.
  B}\ }\textbf {\bibinfo {volume} {27}},\ \bibinfo {pages} {361} (\bibinfo
  {year} {2010})}\BibitemShut {NoStop}%
\bibitem [{\citenamefont {Watkins}\ and\ \citenamefont
  {Kuzyk}(2011)}]{watkins11.01}%
  \BibitemOpen
  \bibfield  {author} {\bibinfo {author} {\bibfnamefont {D.~S.}\ \bibnamefont
  {Watkins}}\ and\ \bibinfo {author} {\bibfnamefont {M.~G.}\ \bibnamefont
  {Kuzyk}},\ }\href@noop {} {\bibfield  {journal} {\bibinfo  {journal} {J.
  Chem. Phys.}\ }\textbf {\bibinfo {volume} {134}},\ \bibinfo {pages} {094109}
  (\bibinfo {year} {2011})}\BibitemShut {NoStop}%
\bibitem [{\citenamefont {Atherton}\ \emph {et~al.}(2012)\citenamefont
  {Atherton}, \citenamefont {Lesnefsky}, \citenamefont {Wiggers},\ and\
  \citenamefont {Petschek}}]{ather12.01}%
  \BibitemOpen
  \bibfield  {author} {\bibinfo {author} {\bibfnamefont {T.~J.}\ \bibnamefont
  {Atherton}}, \bibinfo {author} {\bibfnamefont {J.}~\bibnamefont {Lesnefsky}},
  \bibinfo {author} {\bibfnamefont {G.~A.}\ \bibnamefont {Wiggers}}, \ and\
  \bibinfo {author} {\bibfnamefont {R.~G.}\ \bibnamefont {Petschek}},\
  }\href@noop {} {\bibfield  {journal} {\bibinfo  {journal} {J. Opt. Soc. Am.
  B}\ }\textbf {\bibinfo {volume} {29}},\ \bibinfo {pages} {513} (\bibinfo
  {year} {2012})}\BibitemShut {NoStop}%
\bibitem [{\citenamefont {Burke}\ \emph {et~al.}(2013)\citenamefont {Burke},
  \citenamefont {Atherton}, \citenamefont {Lesnefsky},\ and\ \citenamefont
  {Petschek}}]{burke13.01}%
  \BibitemOpen
  \bibfield  {author} {\bibinfo {author} {\bibfnamefont {C.~J.}\ \bibnamefont
  {Burke}}, \bibinfo {author} {\bibfnamefont {T.~J.}\ \bibnamefont {Atherton}},
  \bibinfo {author} {\bibfnamefont {J.}~\bibnamefont {Lesnefsky}}, \ and\
  \bibinfo {author} {\bibfnamefont {R.~G.}\ \bibnamefont {Petschek}},\
  }\href@noop {} {\bibfield  {journal} {\bibinfo  {journal} {J. Opt. Soc. Am.
  B}\ }\textbf {\bibinfo {volume} {30}},\ \bibinfo {pages} {1438} (\bibinfo
  {year} {2013})}\BibitemShut {NoStop}%
\bibitem [{\citenamefont {Cole}\ \emph {et~al.}(2002)\citenamefont {Cole},
  \citenamefont {Copley}, \citenamefont {McIntyre}, \citenamefont {Howard},
  \citenamefont {Szablewski},\ and\ \citenamefont {Cross}}]{cole02.01}%
  \BibitemOpen
  \bibfield  {author} {\bibinfo {author} {\bibfnamefont {J.~M.}\ \bibnamefont
  {Cole}}, \bibinfo {author} {\bibfnamefont {R.~C.~B.}\ \bibnamefont {Copley}},
  \bibinfo {author} {\bibfnamefont {G.~J.}\ \bibnamefont {McIntyre}}, \bibinfo
  {author} {\bibfnamefont {J.~A.~K.}\ \bibnamefont {Howard}}, \bibinfo {author}
  {\bibfnamefont {M.}~\bibnamefont {Szablewski}}, \ and\ \bibinfo {author}
  {\bibfnamefont {G.~H.}\ \bibnamefont {Cross}},\ }\href@noop {} {\bibfield
  {journal} {\bibinfo  {journal} {Phys. Rev. B}\ }\textbf {\bibinfo {volume}
  {65}},\ \bibinfo {pages} {125107} (\bibinfo {year} {2002})}\BibitemShut
  {NoStop}%
\bibitem [{\citenamefont {Cole}(2003)}]{cole03.01}%
  \BibitemOpen
  \bibfield  {author} {\bibinfo {author} {\bibfnamefont {J.~M.}\ \bibnamefont
  {Cole}},\ }\href@noop {} {\bibfield  {journal} {\bibinfo  {journal} {Phil.
  Trans. R. Soc. Lond.}\ }\textbf {\bibinfo {volume} {361}},\ \bibinfo {pages}
  {2751} (\bibinfo {year} {2003})}\BibitemShut {NoStop}%
\bibitem [{\citenamefont {Higginbotham}\ \emph {et~al.}(1993)\citenamefont
  {Higginbotham}, \citenamefont {Cole}, \citenamefont {Blood-Forsythe},\ and\
  \citenamefont {Hickstein}}]{higgi12.01}%
  \BibitemOpen
  \bibfield  {author} {\bibinfo {author} {\bibfnamefont {A.~P.}\ \bibnamefont
  {Higginbotham}}, \bibinfo {author} {\bibfnamefont {J.~M.}\ \bibnamefont
  {Cole}}, \bibinfo {author} {\bibfnamefont {M.~A.}\ \bibnamefont
  {Blood-Forsythe}}, \ and\ \bibinfo {author} {\bibfnamefont {D.~D.}\
  \bibnamefont {Hickstein}},\ }\href@noop {} {\bibfield  {journal} {\bibinfo
  {journal} {J. Appl. Phys.}\ }\textbf {\bibinfo {volume} {111}},\ \bibinfo
  {pages} {033512} (\bibinfo {year} {1993})}\BibitemShut {NoStop}%
\bibitem [{\citenamefont {Shafei}\ \emph {et~al.}(2012)\citenamefont {Shafei},
  \citenamefont {Lytel},\ and\ \citenamefont {Kuzyk}}]{shafe12.01}%
  \BibitemOpen
  \bibfield  {author} {\bibinfo {author} {\bibfnamefont {S.}~\bibnamefont
  {Shafei}}, \bibinfo {author} {\bibfnamefont {R.}~\bibnamefont {Lytel}}, \
  and\ \bibinfo {author} {\bibfnamefont {M.~G.}\ \bibnamefont {Kuzyk}},\
  }\href@noop {} {\bibfield  {journal} {\bibinfo  {journal} {J. Opt. Soc. Am.}\
  }\textbf {\bibinfo {volume} {29}},\ \bibinfo {pages} {3419} (\bibinfo {year}
  {2012})}\BibitemShut {NoStop}%
\bibitem [{\citenamefont {Lytel}\ and\ \citenamefont
  {Kuzyk}(2013)}]{lytel13.01}%
  \BibitemOpen
  \bibfield  {author} {\bibinfo {author} {\bibfnamefont {R.}~\bibnamefont
  {Lytel}}\ and\ \bibinfo {author} {\bibfnamefont {M.~G.}\ \bibnamefont
  {Kuzyk}},\ }\href@noop {} {\bibfield  {journal} {\bibinfo  {journal} {J.
  Nonlinear Optic. Phys. Mat.}\ }\textbf {\bibinfo {volume} {22}},\ \bibinfo
  {pages} {1350041} (\bibinfo {year} {2013})}\BibitemShut {NoStop}%
\bibitem [{\citenamefont {Lytel}\ \emph {et~al.}(2013)\citenamefont {Lytel},
  \citenamefont {Shafei}, \citenamefont {Smith},\ and\ \citenamefont
  {Kuzyk}}]{lytel13.02}%
  \BibitemOpen
  \bibfield  {author} {\bibinfo {author} {\bibfnamefont {R.}~\bibnamefont
  {Lytel}}, \bibinfo {author} {\bibfnamefont {S.}~\bibnamefont {Shafei}},
  \bibinfo {author} {\bibfnamefont {J.~H.}\ \bibnamefont {Smith}}, \ and\
  \bibinfo {author} {\bibfnamefont {M.~G.}\ \bibnamefont {Kuzyk}},\ }\href@noop
  {} {\bibfield  {journal} {\bibinfo  {journal} {Phys. Rev. A}\ }\textbf
  {\bibinfo {volume} {87}},\ \bibinfo {pages} {043824} (\bibinfo {year}
  {2013})}\BibitemShut {NoStop}%
\bibitem [{\citenamefont {Orr}\ and\ \citenamefont {Ward}(1971)}]{orr71.01}%
  \BibitemOpen
  \bibfield  {author} {\bibinfo {author} {\bibfnamefont {B.~J.}\ \bibnamefont
  {Orr}}\ and\ \bibinfo {author} {\bibfnamefont {J.~F.}\ \bibnamefont {Ward}},\
  }\href@noop {} {\bibfield  {journal} {\bibinfo  {journal} {Molecular
  Physics}\ }\textbf {\bibinfo {volume} {20}},\ \bibinfo {pages} {513}
  (\bibinfo {year} {1971})}\BibitemShut {NoStop}%
\bibitem [{\citenamefont {Levinger}\ and\ \citenamefont
  {Bethea}(1957)}]{levin57.01}%
  \BibitemOpen
  \bibfield  {author} {\bibinfo {author} {\bibfnamefont {J.~S.}\ \bibnamefont
  {Levinger}}\ and\ \bibinfo {author} {\bibfnamefont {C.~G.}\ \bibnamefont
  {Bethea}},\ }\href@noop {} {\bibfield  {journal} {\bibinfo  {journal} {Phys.
  Rev.}\ }\textbf {\bibinfo {volume} {106}},\ \bibinfo {pages} {1191} (\bibinfo
  {year} {1957})}\BibitemShut {NoStop}%
\bibitem [{\citenamefont {Leung}\ \emph {et~al.}(1986)\citenamefont {Leung},
  \citenamefont {Rustgi},\ and\ \citenamefont {Long}}]{leung86.01}%
  \BibitemOpen
  \bibfield  {author} {\bibinfo {author} {\bibfnamefont {P.~T.}\ \bibnamefont
  {Leung}}, \bibinfo {author} {\bibfnamefont {M.~L.}\ \bibnamefont {Rustgi}}, \
  and\ \bibinfo {author} {\bibfnamefont {S.~A.~T.}\ \bibnamefont {Long}},\
  }\href@noop {} {\bibfield  {journal} {\bibinfo  {journal} {Phys. Rev. A}\
  }\textbf {\bibinfo {volume} {33}},\ \bibinfo {pages} {2827} (\bibinfo {year}
  {1986})}\BibitemShut {NoStop}%
\bibitem [{\citenamefont {Cohen}\ and\ \citenamefont
  {Leung}(1998)}]{cohen98.01}%
  \BibitemOpen
  \bibfield  {author} {\bibinfo {author} {\bibfnamefont {S.~M.}\ \bibnamefont
  {Cohen}}\ and\ \bibinfo {author} {\bibfnamefont {P.~T.}\ \bibnamefont
  {Leung}},\ }\href@noop {} {\bibfield  {journal} {\bibinfo  {journal} {Phys.
  Rev. A}\ }\textbf {\bibinfo {volume} {57}},\ \bibinfo {pages} {4994}
  (\bibinfo {year} {1998})}\BibitemShut {NoStop}%
\bibitem [{\citenamefont {Sinky}\ and\ \citenamefont
  {Leung}(2006)}]{sinky06.01}%
  \BibitemOpen
  \bibfield  {author} {\bibinfo {author} {\bibfnamefont {H.}~\bibnamefont
  {Sinky}}\ and\ \bibinfo {author} {\bibfnamefont {P.~T.}\ \bibnamefont
  {Leung}},\ }\href@noop {} {\bibfield  {journal} {\bibinfo  {journal} {Phys.
  Rev. A}\ }\textbf {\bibinfo {volume} {74}},\ \bibinfo {pages} {034703}
  (\bibinfo {year} {2006})}\BibitemShut {NoStop}%
\bibitem [{\citenamefont {Foldy}\ and\ \citenamefont
  {Wouthuysen}(1950)}]{foldy50.01}%
  \BibitemOpen
  \bibfield  {author} {\bibinfo {author} {\bibfnamefont {L.~L.}\ \bibnamefont
  {Foldy}}\ and\ \bibinfo {author} {\bibfnamefont {S.~A.}\ \bibnamefont
  {Wouthuysen}},\ }\href@noop {} {\bibfield  {journal} {\bibinfo  {journal} {J.
  Opt. Soc. Am. B}\ }\textbf {\bibinfo {volume} {78}},\ \bibinfo {pages} {29}
  (\bibinfo {year} {1950})}\BibitemShut {NoStop}%
\bibitem [{\citenamefont {Greiner}(2000)}]{grein00.01}%
  \BibitemOpen
  \bibfield  {author} {\bibinfo {author} {\bibfnamefont {W.}~\bibnamefont
  {Greiner}},\ }\href@noop {} {\emph {\bibinfo {title} {Relativistic quantum
  mechanics. Wave equations}}},\ \bibinfo {edition} {3rd}\ ed.\ (\bibinfo
  {publisher} {Springer},\ \bibinfo {address} {New York},\ \bibinfo {year}
  {2000})\BibitemShut {NoStop}%
\bibitem [{\citenamefont {Douglas}\ and\ \citenamefont
  {Kroll}(1974)}]{dougl74.01}%
  \BibitemOpen
  \bibfield  {author} {\bibinfo {author} {\bibfnamefont {M.}~\bibnamefont
  {Douglas}}\ and\ \bibinfo {author} {\bibfnamefont {N.~M.}\ \bibnamefont
  {Kroll}},\ }\href@noop {} {\bibfield  {journal} {\bibinfo  {journal} {Ann.
  Phys.}\ }\textbf {\bibinfo {volume} {82}},\ \bibinfo {pages} {89} (\bibinfo
  {year} {1974})}\BibitemShut {NoStop}%
\bibitem [{\citenamefont {Hess}(1986)}]{hess86.01}%
  \BibitemOpen
  \bibfield  {author} {\bibinfo {author} {\bibfnamefont {B.~A.}\ \bibnamefont
  {Hess}},\ }\href@noop {} {\bibfield  {journal} {\bibinfo  {journal} {Phys.
  Rev. A}\ }\textbf {\bibinfo {volume} {33}},\ \bibinfo {pages} {3742}
  (\bibinfo {year} {1986})}\BibitemShut {NoStop}%
\bibitem [{\citenamefont {Nakajima}\ and\ \citenamefont
  {Hirao}(2000)}]{nakaj00.01}%
  \BibitemOpen
  \bibfield  {author} {\bibinfo {author} {\bibfnamefont {T.}~\bibnamefont
  {Nakajima}}\ and\ \bibinfo {author} {\bibfnamefont {K.}~\bibnamefont
  {Hirao}},\ }\href@noop {} {\bibfield  {journal} {\bibinfo  {journal} {J.
  Chem. Phys.}\ }\textbf {\bibinfo {volume} {113}},\ \bibinfo {pages} {7786}
  (\bibinfo {year} {2000})}\BibitemShut {NoStop}%
\bibitem [{\citenamefont {Reiher}(2012)}]{reihe12.01}%
  \BibitemOpen
  \bibfield  {author} {\bibinfo {author} {\bibfnamefont {M.}~\bibnamefont
  {Reiher}},\ }\href@noop {} {\bibfield  {journal} {\bibinfo  {journal} {WIREs
  Comput. Mol. Sci.}\ }\textbf {\bibinfo {volume} {2}},\ \bibinfo {pages} {139}
  (\bibinfo {year} {2012})}\BibitemShut {NoStop}%
\bibitem [{\citenamefont {van Lenthe}\ \emph {et~al.}(1993)\citenamefont {van
  Lenthe}, \citenamefont {Baerends},\ and\ \citenamefont
  {Snijders}}]{lenth93.01}%
  \BibitemOpen
  \bibfield  {author} {\bibinfo {author} {\bibfnamefont {E.}~\bibnamefont {van
  Lenthe}}, \bibinfo {author} {\bibfnamefont {E.~J.}\ \bibnamefont {Baerends}},
  \ and\ \bibinfo {author} {\bibfnamefont {J.~G.}\ \bibnamefont {Snijders}},\
  }\href@noop {} {\bibfield  {journal} {\bibinfo  {journal} {J. Chem. Phys.}\
  }\textbf {\bibinfo {volume} {99}},\ \bibinfo {pages} {4597} (\bibinfo {year}
  {1993})}\BibitemShut {NoStop}%
\bibitem [{\citenamefont {van Lenthe}\ \emph {et~al.}(1994)\citenamefont {van
  Lenthe}, \citenamefont {Baerends},\ and\ \citenamefont
  {Snijders}}]{lenth94.01}%
  \BibitemOpen
  \bibfield  {author} {\bibinfo {author} {\bibfnamefont {E.}~\bibnamefont {van
  Lenthe}}, \bibinfo {author} {\bibfnamefont {E.~J.}\ \bibnamefont {Baerends}},
  \ and\ \bibinfo {author} {\bibfnamefont {J.~G.}\ \bibnamefont {Snijders}},\
  }\href@noop {} {\bibfield  {journal} {\bibinfo  {journal} {J. Chem. Phys.}\
  }\textbf {\bibinfo {volume} {101}},\ \bibinfo {pages} {9783} (\bibinfo {year}
  {1994})}\BibitemShut {NoStop}%
\bibitem [{\citenamefont {van Lenthe}\ \emph {et~al.}(1996)\citenamefont {van
  Lenthe}, \citenamefont {Leeuwen}, \citenamefont {Baerends},\ and\
  \citenamefont {Snijders}}]{lenth96.01}%
  \BibitemOpen
  \bibfield  {author} {\bibinfo {author} {\bibfnamefont {E.}~\bibnamefont {van
  Lenthe}}, \bibinfo {author} {\bibfnamefont {R.~v.}\ \bibnamefont {Leeuwen}},
  \bibinfo {author} {\bibfnamefont {E.~J.}\ \bibnamefont {Baerends}}, \ and\
  \bibinfo {author} {\bibfnamefont {J.~G.}\ \bibnamefont {Snijders}},\
  }\href@noop {} {\bibfield  {journal} {\bibinfo  {journal} {Int. J. Quant.
  Chem.}\ }\textbf {\bibinfo {volume} {57}},\ \bibinfo {pages} {281} (\bibinfo
  {year} {1996})}\BibitemShut {NoStop}%
\bibitem [{\citenamefont {Filatov}\ and\ \citenamefont
  {Cremer}(2003{\natexlab{a}})}]{filat03.01}%
  \BibitemOpen
  \bibfield  {author} {\bibinfo {author} {\bibfnamefont {M.}~\bibnamefont
  {Filatov}}\ and\ \bibinfo {author} {\bibfnamefont {D.}~\bibnamefont
  {Cremer}},\ }\href@noop {} {\bibfield  {journal} {\bibinfo  {journal} {J.
  Chem. Phys.}\ }\textbf {\bibinfo {volume} {118}},\ \bibinfo {pages} {6741}
  (\bibinfo {year} {2003}{\natexlab{a}})}\BibitemShut {NoStop}%
\bibitem [{\citenamefont {Filatov}\ and\ \citenamefont
  {Cremer}(2003{\natexlab{b}})}]{filat03.02}%
  \BibitemOpen
  \bibfield  {author} {\bibinfo {author} {\bibfnamefont {M.}~\bibnamefont
  {Filatov}}\ and\ \bibinfo {author} {\bibfnamefont {D.}~\bibnamefont
  {Cremer}},\ }\href@noop {} {\bibfield  {journal} {\bibinfo  {journal} {J.
  Chem. Phys.}\ }\textbf {\bibinfo {volume} {119}},\ \bibinfo {pages} {11526}
  (\bibinfo {year} {2003}{\natexlab{b}})}\BibitemShut {NoStop}%
\bibitem [{\citenamefont {Kutzelnigg}\ and\ \citenamefont
  {Liu}(2005)}]{kutze05.01}%
  \BibitemOpen
  \bibfield  {author} {\bibinfo {author} {\bibfnamefont {W.}~\bibnamefont
  {Kutzelnigg}}\ and\ \bibinfo {author} {\bibfnamefont {W.}~\bibnamefont
  {Liu}},\ }\href@noop {} {\bibfield  {journal} {\bibinfo  {journal} {J. Chem.
  Phys.}\ }\textbf {\bibinfo {volume} {123}},\ \bibinfo {pages} {241102}
  (\bibinfo {year} {2005})}\BibitemShut {NoStop}%
\bibitem [{\citenamefont {Kutzelnigg}\ and\ \citenamefont
  {Liu}(2006)}]{kutze06.01}%
  \BibitemOpen
  \bibfield  {author} {\bibinfo {author} {\bibfnamefont {W.}~\bibnamefont
  {Kutzelnigg}}\ and\ \bibinfo {author} {\bibfnamefont {W.}~\bibnamefont
  {Liu}},\ }\href@noop {} {\bibfield  {journal} {\bibinfo  {journal} {Mol.
  Phys.}\ }\textbf {\bibinfo {volume} {104}},\ \bibinfo {pages} {2225}
  (\bibinfo {year} {2006})}\BibitemShut {NoStop}%
\bibitem [{\citenamefont {Kutzelnigg}\ and\ \citenamefont
  {Liu}(2007)}]{kutze07.01}%
  \BibitemOpen
  \bibfield  {author} {\bibinfo {author} {\bibfnamefont {W.}~\bibnamefont
  {Kutzelnigg}}\ and\ \bibinfo {author} {\bibfnamefont {W.}~\bibnamefont
  {Liu}},\ }\href@noop {} {\bibfield  {journal} {\bibinfo  {journal} {J. Chem.
  Phys.}\ }\textbf {\bibinfo {volume} {126}},\ \bibinfo {pages} {114107}
  (\bibinfo {year} {2007})}\BibitemShut {NoStop}%
\bibitem [{\citenamefont {Sewell}(1949)}]{sewel49.01}%
  \BibitemOpen
  \bibfield  {author} {\bibinfo {author} {\bibfnamefont {G.~L.}\ \bibnamefont
  {Sewell}},\ }\href@noop {} {\bibfield  {journal} {\bibinfo  {journal} {Math.
  Proc. Cambridge}\ }\textbf {\bibinfo {volume} {45}},\ \bibinfo {pages} {678}
  (\bibinfo {year} {1949})}\BibitemShut {NoStop}%
\bibitem [{\citenamefont {Boyle}\ \emph {et~al.}(1966)\citenamefont {Boyle},
  \citenamefont {Buckingham}, \citenamefont {Disch},\ and\ \citenamefont
  {Dunmur}}]{boyle01.66}%
  \BibitemOpen
  \bibfield  {author} {\bibinfo {author} {\bibfnamefont {L.~L.}\ \bibnamefont
  {Boyle}}, \bibinfo {author} {\bibfnamefont {A.~D.}\ \bibnamefont
  {Buckingham}}, \bibinfo {author} {\bibfnamefont {R.~L.}\ \bibnamefont
  {Disch}}, \ and\ \bibinfo {author} {\bibfnamefont {D.~A.}\ \bibnamefont
  {Dunmur}},\ }\href@noop {} {\bibfield  {journal} {\bibinfo  {journal} {J.
  Chem. Phys.}\ }\textbf {\bibinfo {volume} {45}},\ \bibinfo {pages} {1318}
  (\bibinfo {year} {1966})}\BibitemShut {NoStop}%
\bibitem [{\citenamefont {Shelton}(1987)}]{shelt87.03}%
  \BibitemOpen
  \bibfield  {author} {\bibinfo {author} {\bibfnamefont {D.~P.}\ \bibnamefont
  {Shelton}},\ }\href@noop {} {\bibfield  {journal} {\bibinfo  {journal} {Phys.
  Rev. A}\ }\textbf {\bibinfo {volume} {36}},\ \bibinfo {pages} {3032}
  (\bibinfo {year} {1987})}\BibitemShut {NoStop}%
\bibitem [{\citenamefont {Griffiths}(1995)}]{griff95.01}%
  \BibitemOpen
  \bibfield  {author} {\bibinfo {author} {\bibfnamefont {D.~J.}\ \bibnamefont
  {Griffiths}},\ }\href@noop {} {\emph {\bibinfo {title} {Introduction to
  Quantum Mechanics}}}\ (\bibinfo  {publisher} {Prentice Hall, Inc.},\ \bibinfo
  {address} {Upper Saddle River, NJ},\ \bibinfo {year} {1995})\BibitemShut
  {NoStop}%
\bibitem [{\citenamefont {Bethe}\ and\ \citenamefont
  {Salpeter}(1977)}]{Bethe77.01}%
  \BibitemOpen
  \bibfield  {author} {\bibinfo {author} {\bibfnamefont {H.~A.}\ \bibnamefont
  {Bethe}}\ and\ \bibinfo {author} {\bibfnamefont {E.~E.}\ \bibnamefont
  {Salpeter}},\ }\href@noop {} {\emph {\bibinfo {title} {Quantum Mechanics of
  One and Two Electron Atoms}}}\ (\bibinfo  {publisher} {Plenum},\ \bibinfo
  {address} {New York},\ \bibinfo {year} {1977})\BibitemShut {NoStop}%
\bibitem [{\citenamefont {Garstang}(1971)}]{garst01.71}%
  \BibitemOpen
  \bibfield  {author} {\bibinfo {author} {\bibfnamefont {R.~H.}\ \bibnamefont
  {Garstang}}\ }(\bibinfo  {publisher} {Colorado Associated University Press},\
  \bibinfo {address} {Boulder},\ \bibinfo {year} {1971})\ pp.\ \bibinfo {pages}
  {153--167}\BibitemShut {NoStop}%
\bibitem [{\citenamefont {Younger}\ and\ \citenamefont
  {Weiss}(1975)}]{young75.01}%
  \BibitemOpen
  \bibfield  {author} {\bibinfo {author} {\bibfnamefont {S.~M.}\ \bibnamefont
  {Younger}}\ and\ \bibinfo {author} {\bibfnamefont {A.~W.}\ \bibnamefont
  {Weiss}},\ }\href@noop {} {\bibfield  {journal} {\bibinfo  {journal} {J. Res.
  Nat. Stand. Sec. A}\ }\textbf {\bibinfo {volume} {79A}},\ \bibinfo {pages}
  {629} (\bibinfo {year} {1975})}\BibitemShut {NoStop}%
\end{thebibliography}

%

\end{document}